\documentclass[authoryear,preprint,12pt]{elsarticle}

\usepackage{graphicx}
\usepackage[section]{placeins}
\usepackage{capt-of} 
\usepackage{tabularx}
\usepackage{titlesec}
\titlespacing*{\section}{0pt}{6pt plus 2pt minus 2pt}{4pt plus 1pt minus 1pt}
\titlespacing*{\subsection}{0pt}{5pt plus 1pt minus 1pt}{3pt plus 1pt minus 1pt}
\renewcommand{\arraystretch}{0.8}      
\usepackage{amsmath}

\makeatletter
\g@addto@macro\normalsize{%
  \setlength{\abovedisplayskip}{4pt plus 1pt minus 2pt}%
  \setlength{\belowdisplayskip}{4pt plus 1pt minus 2pt}%
  \setlength{\abovedisplayshortskip}{2pt plus 1pt minus 1pt}%
  \setlength{\belowdisplayshortskip}{2pt plus 1pt minus 1pt}%
  \setlength{\parskip}{0pt}%
}
\makeatother

\usepackage{lineno}       
\usepackage{geometry}     
\usepackage{setspace}     
\usepackage{amsmath, amssymb} 
\usepackage{graphicx}     
\usepackage{booktabs}     
\usepackage{hyperref}     
\modulolinenumbers[5]
\usepackage{amsmath,amsfonts}
\usepackage{algorithmic}
\usepackage{adjustbox} 

\usepackage[linesnumbered,ruled,vlined]{algorithm2e}

\usepackage{array}
\usepackage{xcolor}
\usepackage{float}
\usepackage[caption=false,font=normalsize,labelfont=sf,textfont=sf]{subfig}
\usepackage{textcomp}
\usepackage{stfloats}
\usepackage{soul} 
\usepackage{url}
\usepackage{verbatim}
\usepackage{graphicx}
\usepackage{tabularx,booktabs,textcomp}
\usepackage{amsthm} 
\usepackage{longtable}
\usepackage{tabularray}

\usepackage{amsmath,amssymb,amsthm}

\usepackage{enumitem}
\usepackage{tabularx}

\usepackage{booktabs}
\usepackage{colortbl}  
\journal{Elsevier}

\usepackage{ragged2e}
\justifying
\begin{document}

\begin{frontmatter}

\title{FaultXformer: A Transformer-Encoder Based Fault Classification and Location Identification model in PMU-Integrated Active Electrical Distribution System}



\author[1,2]{Kriti Thakur}

\author[2]{Alivelu Manga Parimi}
\author[1]{Mayukha Pal\corref{mycorrespondingauthor}}
\ead{mayukha.pal@in.abb.com}
\cortext[mycorrespondingauthor]{Corresponding author}


\affiliation[1]{organization={ABB Ability Innovation Center},
     addressline={Asea Brown Boveri Company},
    city={Hyderabad},
    postcode={500084},
    state={Telangana},
    country={India}}

\affiliation[2]{organization={Department of Electrical and Electronics Engineering},
addressline={Birla Institute of Technology and Science, Pilani- Hyderabad Campus},
city={Hyderabad},
 postcode={500078},
state={Telangana},
country={India}} 

\begin{abstract}

Accurate fault detection and localization in electrical distribution systems is crucial, especially with the increasing integration of distributed energy resources (DERs), which inject greater variability and complexity into grid operations. In this study, FaultXformer is proposed, a Transformer encoder-based architecture developed for automatic fault analysis using real-time current data obtained from phasor measurement unit (PMU). The approach utilizes time-series current data to initially extract rich temporal information in stage 1, which is crucial for identifying the fault type and precisely determining its location across multiple nodes. In Stage 2, these extracted features are processed to differentiate among distinct fault types and identify the respective fault location within the distribution system. Thus, this dual-stage transformer encoder pipeline enables high-fidelity representation learning, considerably boosting the performance of the work. The model was validated on a dataset generated from the IEEE 13-node test feeder, simulated with 20 separate fault locations and several DER integration scenarios, utilizing current measurements from four strategically located PMUs. To demonstrate robust performance evaluation, stratified 10-fold cross-validation is performed.  FaultXformer achieved average accuracies of 98.76\% in fault type classification and 98.92\% in fault location identification across cross-validation, consistently surpassing conventional deep learning baselines convolutional neural network (CNN), recurrent neural network (RNN). long short-term memory (LSTM) by 1.70\%, 34.95\%, and 2.04\% in classification accuracy and by 10.82\%, 40.89\%, and 6.27\% in location accuracy, respectively. These results demonstrate the efficacy of the proposed model with significant DER penetration.

\end{abstract}

\end{frontmatter}

\section{Introduction}
\label{section:Introduction}
Electrical distribution systems (EDS), the backbone of modern energy infrastructure that deliver power from substations to end users. It must maintain operational reliability despite increasing demand, as faults from lightning, equipment failures, or disturbances may significantly compromise system performance and safety \cite{DWIVEDI2025110114}.

Faults in the distribution grid lead to power outages, financial losses, and decreased customer satisfaction, so degrading the entire dependability of the system \cite{mirshekali2023deep}. The growing complexity of modern power distribution networks, especially with the incorporation of distributed energy resources (DER) such as solar photovoltaics (PV) and wind turbines, has significantly increased the demand for robust and efficient fault analysis technologies. Routine maintenance and continuous fault monitoring are critical for recognising early degradation, isolating faults, and preventing failures. Effective fault identification and localization save downtime, increase equipment lifespan, and enhance the dependability and resilience of electrical distribution networks.

\subsection{Related Works}

Fault detection and localisation in power distribution systems has been a key research focus due to its importance in ensuring system dependability and operational safety. Conventional fault location approaches could be classified into impedance-based (IB) and traveling-wave (TW)-based methods \cite{stefanidou2022review}. The IB approach uses Kirchhoff’s principles to pre- and post-fault voltage and current measurements with the line impedance matrix to determine the fault distance from the feeder. The TW approach, in contrast, reveals fault location by examining the propagation duration of high-frequency waves created during a fault.  While IB approaches require complicated iterative computations and may yield multiple possible locations \cite{aboshady2019new}, TW methods require high sampling rates, precise synchronization, wide-band communication, and complex system infrastructure \cite{shi2019fault}.

The deployment of smart devices has provided distribution utilities with significantly more data for fault analysis, creating opportunities for knowledge-based techniques to leverage this expanded information for enhanced fault identification. The main machine learning methodologies applicable to the fault location approach include support vector machines (SVM), k-nearest neighbours (KNN), artificial neural networks (ANN), and convolutional neural networks (CNN)\cite{gururajapathy2017fault}. A data-driven strategy employing an ensemble classifier was presented for fault identification in distribution systems using smart meter data in \cite{dutta2023data}.  However, this approach was confined to detection and did not provide fault-locating capabilities. The work in \cite{LIANG2023109290} developed a cost-effective ANN-based data-fusion model that increases fault location accuracy by merging complementing algorithmic outputs.  Similarly, \cite{ghaemi2022accuracy} presented an ensemble learning technique employing SVM, k-NN, and Random Forest inside a stacking framework, albeit its meta-classifier structure requires considerable computational cost, restricting real-time application.

Recent research has proved the effectiveness of  CNNs for fault classification and location in power distribution systems.  \cite{RAI2021106914} employed raw three-phase voltage and current signals for fault classification without feature extraction but lacked fault location capability.  A vanilla CNN in \cite{yoon2022deep} achieved 99\% accuracy in classification and localization but failed to identify fault phase or distance, while other CNN-based approaches depended on substation voltage and spectrogram analysis for offline localization without fault type detection. The CNN–Transformer model in \cite{thomas2023cnn} attained 98.6\% and 98.1\% accuracy for classification and localization, respectively, but did not consider DER integration.  Table~\ref{tab:comparision} highlights previous investigations, showing that few enable simultaneous classification and localization, and most disregard DER effects and measurement noise, limiting generalization.
\subsection{Motivation and Contribution}
Phasor measurement units (PMUs) are becoming crucial in modern power systems because of their capability to provide high-frequency, time-synchronized measurements of voltage and current phasors \cite{phadke2008synchronized}. These qualities provide essential spatiotemporal insights into grid dynamics, particularly under fault conditions. As EDS grow with the rise of DERs, such as solar PV and wind turbines, traditional model-based methodologies often fall short due to the associated nonlinearity and variability. The existing approaches, such as Graph Convolutional Networks (GCNs)\cite{chen2019fault}, CNNs\cite{siddique2024fault}, Capsule CNNs \cite{mirshekali2023deep}, IB methods \cite{yang2024fault}, and TW approaches have demonstrated substantiated results \cite{cavalari2024enhanced}, while each has its own limitations. CNN-based models have limited capacity to capture long-range dependencies and global temporal relationships in time-series data, as their convolutional filters primarily focus on local patterns within a narrow receptive field. This shortcoming may hinder their performance in situations where understanding complicated, non-local, or sequential dependencies is critical.\cite{kiranyaz20211d}.  Capsule CNNs further enhance this cost due to their complicated dynamic routing algorithms.  GCNs, while suited for network-structured data, depend largely on accurate topological information and are sensitive to changes in the grid structure.In contrast, Transformer-based models process raw time-series PMU data directly. Their self-attention mechanism permits effective modeling of long-range temporal relationships while staying highly parallelizable and scalable \cite{vaswani2017attention}. Applications of Transformers in electrical systems have been researched for load forecasting and anomaly detection \cite{9756020,liu2025vision}, but their usage for fault type and location identification remains limited.

To address the challenges of fault classification and localization under DER integration, the proposed FaultXformer utilises high-resolution PMU data to achieve accurate, robust, and generalizable fault identification across diverse fault types and locations, outperforming existing methods in both precision and computational efficiency.
The novelty of this research lies in the domain-specific adaptation and validation of a Transformer encoder-based framework for PMU-driven fault detection and localization in distribution networks an area where such architectures remain largely unexplored, capturing the unique temporal–spatial dynamics of current signals and establishing a new benchmark for intelligent, data-driven protection in medium-voltage EDS.
\begin{table}[H]
\centering
\caption{Fault Location methods' comparative table.}
\label{tab:comparision}
\begin{adjustbox}{width=\textwidth,center}
\renewcommand{\arraystretch}{0.7}
\setlength{\tabcolsep}{3pt}
\tiny

\begin{tabular}{|p{1.4cm}|p{0.6cm}|p{2.5cm}|p{2cm}|p{0.6cm}|p{2cm}|p{2cm}|}
\hline
\textbf{Reference} & \textbf{PMU considered} & \textbf{Data Type} & \textbf{Noise} & \textbf{DER} & \textbf{Results} & \textbf{Method} \\
\hline

\hline
\cite{cavalari2024enhanced} & Yes & $V, I$ at smart meters or micro PMUs & No & Yes & Error = 5\% & Impedance \\
\hline
\cite{ghaemi2022accuracy} & No & $V, I$ at end branches & Gaussian noise & Yes & FTC = 99\%, FLI = 97\% & Stacked Ensemble Learning \\
\hline
\cite{lv2024improved} & No & $V$ at substation & Gaussian white noise & Yes & Nearby Fault location accuracy = 98\% & Bayesian learning \\
\hline
\cite{mirshekali2023deep} & No & $V$ at substation & Noise & No & 99.9\% FSI accuracy and 0.26\% EFL error & Capsule CNN \\
\hline
\cite{siddique2024fault} & Yes & $I$ from PMUs & No & Yes & 99.46\% in FTP, 0.9\% error in EFL & CNN \\
\hline
\cite{LIANG2023109290} & No & $V, I$ at substation & Yes & Yes & Error = 1.27\% & Hybrid (Impedance + ANN) \\
\hline
\cite{chen2019fault} & No & $I$ at substation & Yes & No & 99.26\% in EFL & GCN \\
\hline
\textbf{Proposed Method} & Yes & $I$ from PMUs & Gaussian noise (1\%–3\%) & Yes (PV and Wind) & 99.56\% in FTC, 99.74\% in FLI & FaultXformer \\
\hline
\end{tabular}
\end{adjustbox}
\vspace{0.4mm}
\tiny
\begin{flushleft}
\textbf{Notation:} CNN — Convolutional Neural Network, ANN — Artificial Neural Network, GCN — Graph Convolutional Network, \\
FTC — Fault Type Classification, FLI — Fault Location Identification, EFL — Exact Fault Location, \\
FSI — Fault Section Identification, FTP — Fault Type and Phase.
\end{flushleft}
\end{table}

The significant contributions of this research include:

\begin{enumerate}
    \item Dual-Stage Transformer Encoder Framework: The proposed dual-stage architecture utilises Transformer Encoders for hierarchical feature extraction and task-specific learning. The initial stage captures rich temporal-spatial representations from raw PMU current data, while the second stage (FaultXformer) comprises two independently optimised Transformer Encoder models focused on fault type classification and fault location identification, respectively. This decoupled design improves accuracy and robustness by enabling each model to concentrate on features that are most relevant to its individual purpose.
    \item Cross-Validated Robustness and Classwise Analysis:
 The models are thoroughly evaluated using 10-fold stratified cross-validation on PMU current signals to guarantee a balanced representation of fault types and locations, with consistently high accuracy, precision, recall, and F1 scores confirming methodological robustness, while classwise analysis emphasises performance variations across common and rare faults to guide targeted feature enhancement and model optimization.
 \item Robustness Under Noise and DER Variations:
 FaultXformer presents sustained accuracy and F1 scores for both fault type classification and fault location identification at Gaussian noise levels of 1–3\% and variable DER penetration.  The continuously excellent performance, exceeding 98\% for classification and 97\% for localization, illustrates the model’s tolerance to signal distortion and operational uncertainty in DER-integrated distribution networks.

\end{enumerate}

The remainder of this manuscript is structured as follows.  Section \ref{section:DN} introduces the test distribution network and PMU data collecting infrastructure. Section \ref{section:phasor} presents fault characteristics using current phasor behavior. Section \ref{section:method} explains the suggested technique, including data preprocessing, feature extraction, model architecture, and training process. Section \ref{section:results} presents results and performance evaluation, while Section \ref{section:conclusion} closes with key findings and potential future research areas.

\section{Modelling of Distribution Network}
\label{section:DN}

\begin{figure}[h]
    \centering
    \includegraphics[width=9cm]{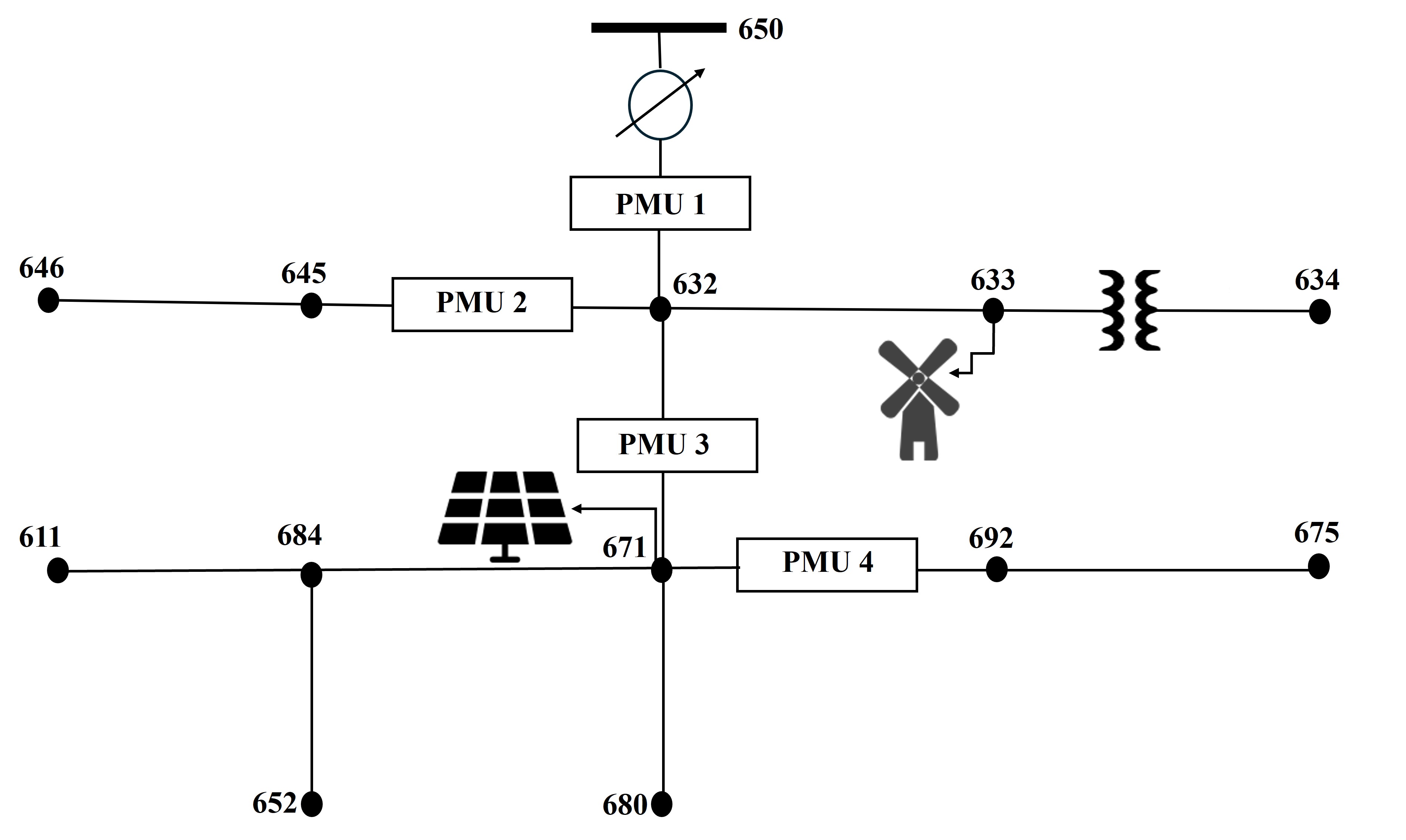}
    \caption{IEEE 13-node test feeder integrated with distributed generation sources.}
    \label{fig:system}
\end{figure}

The proposed fault diagnosis methodology is designed and validated using the IEEE 13-node test feeder \cite{ieee13node}, whose single-line schematic in Figure \ref{fig:system} depicts the DER placements and PMU locations. The simulations were performed using MATLAB 2024 on a workstation equipped with an Intel® Core™ i7-10700 CPU operating at 2.90 GHz and 16 GB of RAM. Operating at 4.16 kV, this benchmark system accurately replicates the complicated characteristics frequently observed in real-world power distribution networks.  It incorporates a variety of system components, including overhead and underground distribution lines with variable layouts, a single substation voltage regulator, multiple load types, and two shunt capacitor banks. Moreover, the feeder includes three-phase, two-phase, and single-phase laterals, thereby providing a comprehensive and realistic testing environment to evaluate fault identification performance under practical distribution network conditions.

To study the impact of integrating DERs, a 500 kW wind energy source and a 300 kW photovoltaic (PV) system have been installed at buses 633 and 671, respectively, in accordance with the principles stated in \cite{abd2015impact}. To provide real-time monitoring and enhance grid observability, PMUs have been placed at strategically chosen locations throughout the distribution network. PMUs enable high-speed, time-synchronized measurements of voltage and current phasors across different locations in the grid, giving complete system visibility. Their placement is directed by network design and identification of essential nodes to enable wide-area coverage while minimizing installation costs. Further, the arrangement is designed to maximize measurement redundancy, thus maintaining full observability even in the event of communication failures or PMU data loss, as mentioned in \cite{shafiullah2018improved}. These PMUs give real-time three-phase current, voltage, and phase measurements, supporting accurate and rapid fault identification.

\section{Fault Characteristics Based on PMU Current Phasor Analysis}
\label{section:phasor}
In this study, the input features for fault type classification and location identification were generated from PMU data, which provide the magnitude as well as the phase angle of the positive-sequence current phasor in a three-phase distribution network.  The PMU block determines the fundamental-frequency phasor representation of the current as:
\begin{equation}
I_{1} = \frac{1}{3}\left( I_{a} + a I_{b} + a^{2} I_{c} \right), \quad a = e^{j120^{\circ}}
\label{eq:positive_sequence_current}
\end{equation}
where $I_{a}$, $I_{b}$, and $I_{c}$ are the phase currents, and $I_{1}$ represents the positive-sequence current component.The PMU output is consequently stated in polar form as:
\begin{equation}
I_{1} = |I_{1}| e^{j\theta_{1}}
\label{eq:phasor_polar_form}
\end{equation}
where $|I_{1}|$ and $\theta_{1}$ denote the magnitude and phase angle of the current phasor, respectively.

Under normal circumstances, the distribution system remains balanced, and both ($|I_{1}|$) and ($\theta_{1}$) of the positive-sequence current show negligible changes. When a fault occurs, the network impedance changes abruptly, resulting in sudden deviations in both magnitude and phase. These changes depend on the fault type and location.

In single line-to-ground (LG) faults, one phase conducts to ground, creating a sharp increase in $|I_{1}|$ and a significant negative shift in $\theta_{1}$. Double line-to-ground (LLG) faults yield the highest unbalanced current and a significant oscillatory phase shift resulting from two parallel fault travels to the ground. In three-phase (LLLG) faults, all phases undergo a symmetric and large current increase, leading to the highest $|I_{1}|$ with little angular difference.

The temporal evolution of $|I_{1}|$ and $\theta_{1}$ therefore produces a different signature for each fault category. These signatures incorporate the nonlinear connection between system impedance, fault path, and DER-induced dynamic behavior, which makes them extremely ideal for data-driven analysis. Accordingly, the proposed Transformer encoder-based methodology uses these features to efficiently discriminate fault types and reliably identify their locations in DER-integrated distribution networks.

Although fault conditions affect all sequence components, this study employs only the positive-sequence current phasor, as commercial PMUs offer more accurate and dependable estimations of this component. Negative- and zero-sequence measurements are frequently degraded by noise, load imbalance, and measurement uncertainty \cite{phadke2008synchronized}.  The positive-sequence current thus gives a robust representation of network dynamics\cite{zhou2011stepwise}, and the achieved high accuracy in fault classification and localization proves its sufficiency for dependable fault identification in DER-integrated distribution networks.

Traditional impedance- and sequence-based analytical techniques, which are limited by linear and steady-state assumptions \cite{li2021asymmetrical}, fail under the nonlinear, time-varying conditions generated by high DER penetration, in contrast Transformer encoder–based deep learning models effectively capture these dynamic dependencies from PMU current phasor sequences \cite{10918689}. By utilising the temporal dynamics of the positive-sequence current $|I_{1}|$ and $\theta_{1}$, the Transformer-encoder based framework successfully captures the intrinsic magnitude–phase coupling and spatio-temporal evolution of fault transients.  Consequently, it achieves accurate and resilient fault type classification and location identification under different DER operating conditions.

\section{Implementation of FaultXformer}
\label{section:method}

\begin{figure*}[h]
    \centering
    \includegraphics[width=14cm]{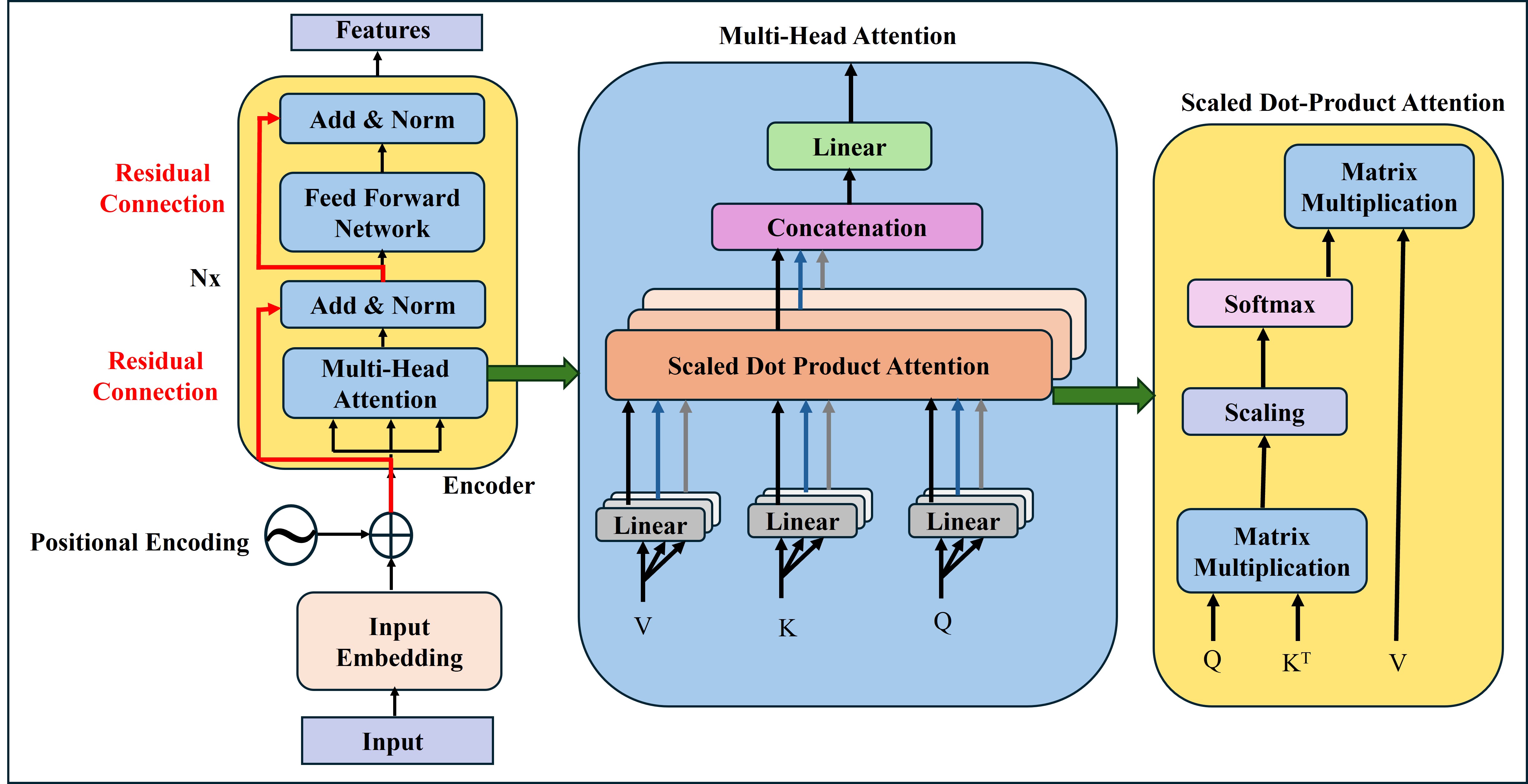}
    \caption{The architectural schematic of a transformer encoder}
    \label{fig:transformer}
\end{figure*}

The Transformer architecture and its advanced versions are fundamentally constructed as a sequence-to-sequence model. It comprises of two basic components: an encoder and a decoder. Each encoder layer comprises a multi-head self-attention mechanism followed by a feed-forward neural network (FNN), with multiple such layers aggregated to constitute the entire encoder. The architectural design of the transformer encoder, incorporating multi-head attention (MHA) with parallel attention layers and a scaled dot-product attention module, is represented in Figure \ref{fig:transformer}. The decoder extends this design by including an encoder–decoder attention module, which runs alongside the self-attention and FNN blocks inherited from the encoder.  These sub-layers are accompanied by residual connections and layer normalization to increase training stability and convergence.  To preserve the temporal structure of the input sequence, the model incorporates positional encodings into the input embeddings.

 The self-attention mechanism acts by substituting each element in the sequence with a weighted sum of all other items, allowing the model to capture relationships throughout the entire sequence. One of the core formulations introduced in \cite{vaswani2017attention} is the \textit{Scaled Dot-Product Attention}, mathematically defined as:
\begin{equation}
\text{Attention}(Q, K, V) = \text{softmax}\left(\frac{QK^\top}{\sqrt{d_k}}\right)V
\label{eq:attention}
\end{equation}
Where \( Q \), \( K \), and \( V \) are the query, key, and value matrices, respectively, and \( d_k \) is the dimensionality of the key vectors.

MHA technique was introduced in  \cite{vaswani2017attention}, allowing the model to concurrently focus on input from many representation subspaces. To enhance representational power, the Transformer uses multiple attention heads operating in parallel, known as the \textit{Multi-Head Attention} mechanism. This is defined as:
\begin{equation}
\text{MultiHead}(Q, K, V) = \text{Concat}(\text{head}_1, \dots, \text{head}_h)W^O
\label{eq:multihead}
\end{equation}
where \(\text{head}_i = \text{Attention}(QW^Q_i, KW^K_i, VW^V_i)\),
\(W^Q_i\), \(W^K_i\), and \(W^V_i\) are projection matrices for queries, keys, and values, and \(W^O\) is the output projection matrix.
Here, \( W^Q_i \), \( W^K_i \), and \( W^V_i \) are learned projection matrices specific to the \( i^{\text{th}} \) attention head, and \( W^O \) is the output projection matrix applied to the concatenated outputs of all heads.

This approach allows the Transformer to record complicated connections from various perspectives across the input sequence, delivering considerable advantages over single-head attention models.

\begin{figure*}[h]
    \centering
    \includegraphics[width=10cm]{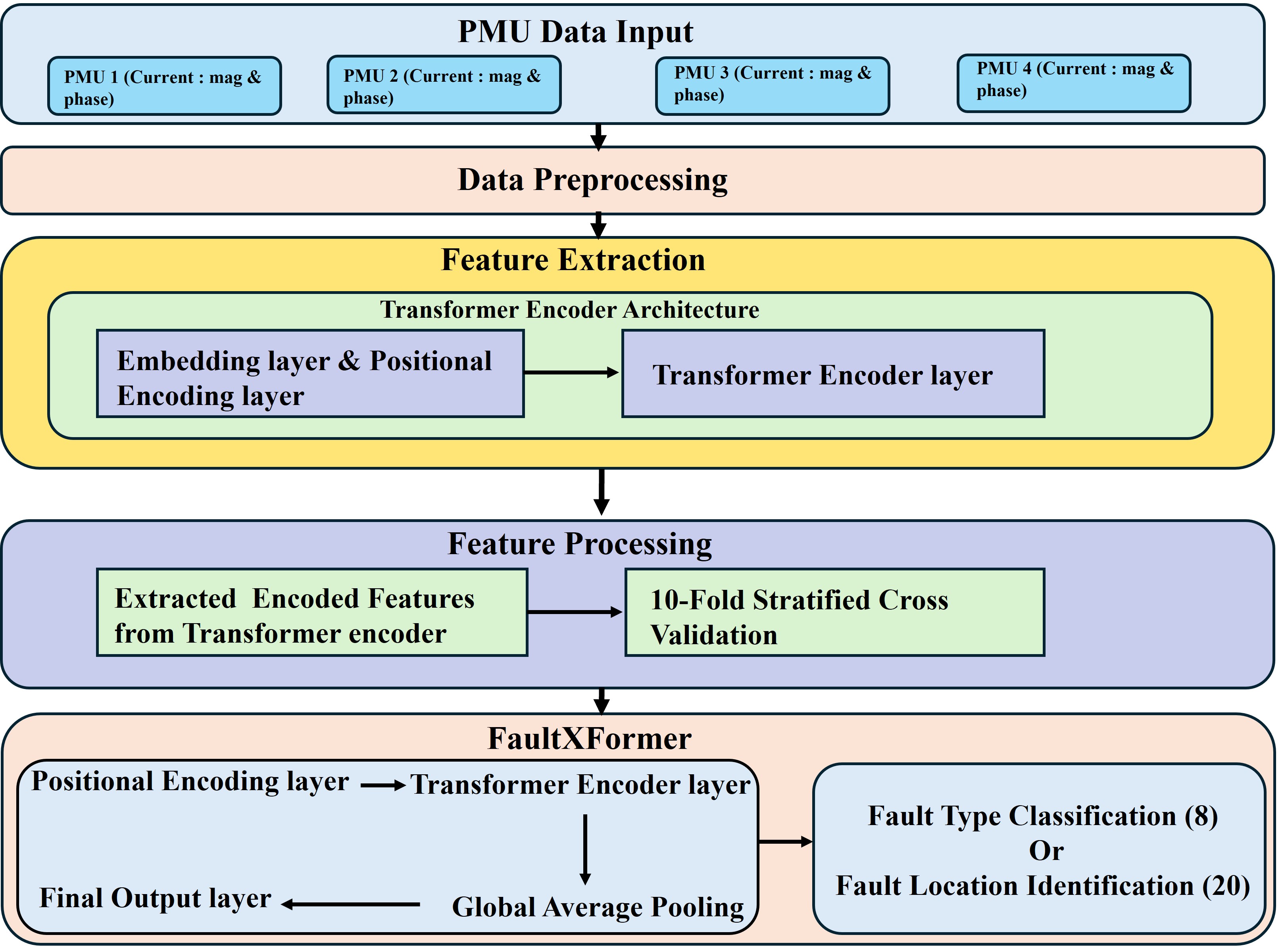}
    \caption{The proposed methodology's process flow diagram for classifying  8 fault types(1 No Fault) and identifying 20 fault locations from PMU data.}
    \label{fig:method}
\end{figure*}

The suggested FaultXformer uses a complete data pipeline for electrical system fault examination. Starting with current magnitude and phase measurements from four PMUs, the system analyses this synchronized data via initial preprocessing stages before feeding it into a specific transformer encoder architecture for feature extraction. This extraction approach comprises embedding layers with positional encoding to capture temporal relationships within the  data.  The retrieved encoded features undergo rigorous validation through a 10-fold stratified cross-validation process to ensure robust performance assessment.  The core FaultXformer module then utilizes these features in a complete transformer encoder-based architecture, comprising positional encoding, transformer encoder layers, and global average pooling, to perform either fault type classification across 8 distinct fault categories(including 1 no fault) or fault location identification across 20 possible locations.  This end-to-end technique efficiently takes advantage of the transformer's attention mechanisms to capture complex spatial and temporal relationships in PMU data, offering high-precision fault detection in EDS.  The workflow diagram is shown in Figure \ref{fig:method} and further explained in subsequent subsections.
\subsection{Data Collection and Data Preprocessing}
PMUs are placed in 4 locations as shown in Figure \ref{fig:system}. The data, i.e., "current" magnitude and phase, are collected from each of the PMUs, which collect 386 samples in 0.1 seconds and are further processed.
Data normalization is a fundamental preprocessing procedure that standardizes PMU signal data for consistent feature representation and enhanced model interpretability.  Z-score normalization is applied individually to the magnitude and phase components of current signals, assuring zero mean and unit variance across features.  This approach mitigates scale disparities, stabilizes Transformer optimization, accelerates convergence, and promotes overall model generalization. The normalization is performed independently on each signal channel using the following equation:
\begin{equation}
z = \frac{x - \mu}{\sigma + \epsilon}
\label{eq:zscore}
\end{equation}
where \( x \) represents the original signal value, \( \mu \) is the mean, \( \sigma \) is the standard deviation of the signal, and \( \epsilon \) is a small constant added (e.g., \( 10^{-8} \)) to ensure numerical stability during division. This normalization is crucial for attention-based models like as Transformers, where input scale significantly impacts dot-product computations. Z-score normalization stabilizes learning dynamics and promotes model generalization across various signal distributions.

In the given approach, time-series data from four PMUs are handled independently rather than concatenated or aggregated.  Each PMU’s current magnitude and phase angle sequences are independently normalized, truncated or padded to a specified length, and fed separately through a Transformer-encoder architecture for feature extraction. This independent processing streamlines the data pipeline, decreases model complexity, and mitigates dangers of overfitting that might result from high-dimensional fused data.  Moreover, it boosts computational efficiency and scalability, making the approach suited for real-time applications. Despite lacking explicit multi-PMU data fusion, the approach attains good performance by leveraging adequate information from individual PMUs, efficiently balancing accuracy and practical implementation. After data preprocessing, the subsequent step is to extract features that will identify significant temporal and spatial patterns from the processed PMU current signals.

\subsection{Transformer Encoder for Feature Extraction and Feature Processing}
In this study, only the encoder component of the Transformer architecture was utilised to extract relevant spatiotemporal characteristics from PMU current signal data.  As depicted in Figure~\ref{fig:transformer}, the input sequence, comprising magnitude and phase data from the PMU, is first sent via an input embedding layer, which projects the raw input into a higher-dimensional feature space designed for transformer processing.  To keep the sequential structure of the time-series data, positional encodings (\ref{eq:pos_enc_sin}) and (\ref{eq:pos_enc_cos}) have been included in the embedded inputs, allowing the model to capture the relative temporal positions of each measurement. 
\begin{equation}
\text{PE}_{(pos, 2i)} = \sin\left(\frac{pos}{10000^{\frac{2i}{d_{\text{model}}}}}\right)
\label{eq:pos_enc_sin}
\end{equation}
\begin{equation}
\text{PE}_{(pos, 2i+1)} = \cos\left(\frac{pos}{10000^{\frac{2i}{d_{\text{model}}}}}\right)
\label{eq:pos_enc_cos}
\end{equation}
where \( pos \) represents the position in the sequence and \( i \) represents the dimension index, \( d_{\text{model}} \) refers to the dimensionality of the embedding space.It denotes the size of the vector employed to encode every position inside the input sequence.

The encoder itself is built of numerous identical layers, each comprising a multi-head self-attention mechanism, then followed by a feed-forward neural network (FNN).  Both components are enclosed with residual connections and layer normalization, which help stabilize training and preserve gradient flow across layers.  The self-attention method allows the model to attend to different time steps simultaneously, learning global dependencies throughout the whole input sequence, which is particularly important for identifying and localizing transient disturbances in the current waveform. The final output of the encoder provides a comprehensive set of features that capture both local and global properties of the fault event, and these features are subsequently exploited by FaultXformer for fault type classification and fault location identification.  The flexible and attention-based structure of the encoder makes it extremely suitable for processing high-resolution PMU data under varied fault scenarios and network configurations. The extracted features are subsequently organised for further processing to enhance their structure and the suitability for the subsequent learning tasks. The encoded features collected from the Transformer encoder that utilizes the whole dataset are separated into ten stratified folds to ensure class balance and then input into the FaultXformer model for training and evaluation.
\subsection{FaultXformer}
FaultXformer is a neural network module developed for sequential data processing using a transformer encoder architecture with positional encoding.  It initially adds positional encoding to the input sequence that helps the network understand the temporal order of the data, which is essential because transformer designs don't automatically capture sequential information.  The basis of FaultXformer is a Transformer Encoder consisting of many encoder layers, each including multi-head self-attention mechanisms and feed-forward networks that allow the model to learn complicated patterns and relationships over the whole sequence \cite{vaswani2017attention}.  The MHA enables the model to attend to distinct parts of the input sequence simultaneously from multiple representation subspaces, capturing various types of relationships. After processing through the transformer encoder, FaultXformer performs global average pooling throughout the time dimension to construct a fixed-size representation independent of input sequence length, which is then sent through a final linear layer to output class predictions. This design is particularly suitable for our work as it can capture both local and global temporal trends in faults while preserving computational efficiency through the transformer's parallel processing capabilities.

\subsection{Evaluation Measures}
The model’s effectiveness in fault classification and location tasks is tested using basic machine learning metrics such as precision, recall, and F1-score, which are derived from the confusion matrix. Table \ref{tab:metrics} illustrates the mathematical definitions of these evaluation metrics.
\begin{table}[H]
\centering
\caption{ Various performance metrics and their mathematical definitions.}
\label{tab:metrics}
\renewcommand{\arraystretch}{0.7}
\setlength{\tabcolsep}{3pt}
\tiny
\begin{tabular}{|l|c|c|}
\hline
\textbf{Evaluation Metric} &\textbf{Definition} &\textbf{Formula} \\
\hline

Precision & Proportion of accurately predicted positive cases compared to the total predicted positives. &$\frac{TP}{TP + FP}$ \\
\hline
Accuracy & evaluates the overall correctness of predictions. & $\frac{TP + TN}{TP + FP + TN + FN}$ \\
\hline
Recall & Ratio of accurately predicted positives to total positives.&$\frac{TP}{TP + FN}$ \\
\hline
F1-Score &  Harmonic mean of precision and recall. &$2 \times \frac{\text{Precision} \times \text{Recall}}{\text{Precision} + \text{Recall}}$ \\
\hline
\end{tabular}

\vspace{0.3cm}
\tiny

\textbf{Notation:} TP — True Positive, TN — True Negative, FP — False Positive, FN — False Negative.
\end{table}
\section{ Results and Discussion}
\label{section:results}
\begin{figure}[H]
\centering
\begin{minipage}{0.48\textwidth}
    \centering
    \includegraphics[width=\linewidth]{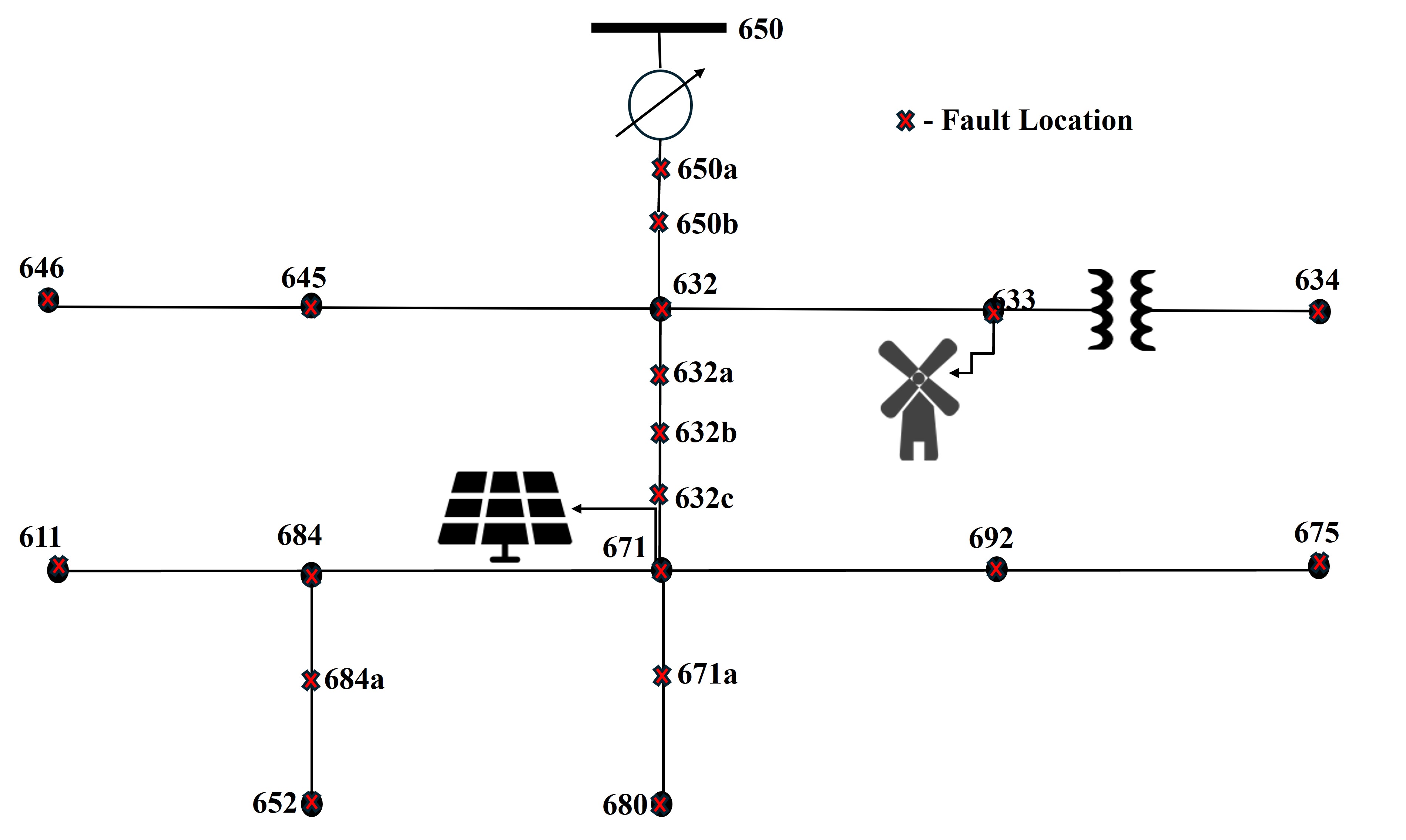}
    \caption{Various fault locations within the IEEE 13-node test feeder.}
    \label{fig:fl}
\end{minipage}
\hfill
\begin{minipage}{0.48\textwidth}
    \centering
     \captionof{table}{Bus Distance from Substation and Assigned Fault Location Labels.}
    \label{tab:node_fault_locations}
    \renewcommand{\arraystretch}{0.7}
    \setlength{\tabcolsep}{3pt}
    \tiny
    \begin{tabular}{|c|c|c|}
        \hline
        \textbf{Bus} & \textbf{Distance (ft)} & \textbf{Fault Label} \\
        \hline
        646 & 2800 & F1 \\
        645 & 2500 & F2 \\
        632 & 2000 & F3 \\
        633 & 2500 & F4 \\
        634 & 2500 & F5 \\
        611 & 4600 & F6 \\
        684 & 4300 & F7 \\
        671 & 4000 & F8 \\
        692 & 4000 & F9 \\
        675 & 4500 & F10 \\
        652 & 5100 & F11 \\
        680 & 5000 & F12 \\
        650 & 0 & F13 \\
        650a & 500 & F14 \\
        650b & 1000 & F15 \\
        632a & 2500 & F16 \\
        632b & 3000 & F17 \\
        632c & 3500 & F18 \\
        684a & 4500 & F19 \\
        671a & 4800 & F20 \\
        \hline
    \end{tabular}
\end{minipage}
\end{figure}

This study examines 8 specific fault types: single-line-to-ground faults (AG, BG, and CG) and double-line-to-ground faults (ABG, BCG, and ACG), triple line to ground fault (ABCG) and 1 no fault, implemented at 20 randomly chosen locations, as depicted in Figure \ref{fig:fl} and the distance from the substation for each location is presented in Table \ref{tab:node_fault_locations}. A systematic sample technique was employed to encompass a broad spectrum of fault scenarios spanning diverse fault locations, resistance levels, and incidence angles. Multiple fault scenarios are simulated to encompass every possible location within the network. The training dataset incorporates a range of fault resistances (0.01, 0.10, 1, 5, 10, 20, and 40 $\Omega$) and various fault inception angles (0°, 30°, 60°, 90°, 120°, 150°, and 180°). The current waveforms and associated phase information constitute time-series data, covering the duration from 0.1 seconds to 0.3 seconds of the simulation. A fault duration of 100 milliseconds is considered uniformly across all cases, occurring between 0.16 seconds and 0.26 seconds.

The hyperparameters presented in Tables \ref{tab:hyperparametersfe} and \ref{tab:hyperparameters} were established by a validation-based grid search to guarantee optimal model performance and stability. Candidate hyperparameters such as learning rate (1e 5 to 1e 3), batch size (16 to 64), attention heads (2 to 6), and embedding dimensions (64 to 128) were systematically optimised based on a 70–20–10 training–validation–test split, with the configuration that achieved the highest validation F1-score selected as the final configuration; the resultant lightweight FaultXformer architecture, detailed in Tables~\ref{tab:hyperparameters} and~\ref{tab:architecture}, minimizes computational overhead while maintaining high efficiency for practical deployment.
The hyperparameters in Table~\ref{tab:hyperparametersfe} significantly govern feature extraction and learning in FaultXformer, utilising current magnitude and phase from all PMUs as inputs. Embedding dimensions of 68 and 90, along with attention heads of 4 and 5, are employed for fault type and location tasks, respectively, ensuring a balanced complexity and efficiency through two encoder layers.  The model, utilising the Adam optimiser with a learning rate of 0.001, a batch size of 32, and cross-entropy loss over 200 and 250 epochs, attained consistent convergence and robust generalisation across both tasks.
\begin{table}[H]
\centering
\begin{minipage}{0.48\textwidth}
\centering
\caption{Hyperparameters for feature extraction for fault type classification and fault location identification}
\label{tab:hyperparametersfe}
\renewcommand{\arraystretch}{0.7}
\setlength{\tabcolsep}{3pt}
\tiny
\begin{tabular}{|p{4cm}|p{1.5cm}|p{1.5cm}|}
\hline
\textbf{Hyperparameters} & \textbf{Fault Type classification} & \textbf{Fault location identification} \\
\hline
Number of input features & 2 & 2 \\
\hline
Embedding dimension for input projection & 68 & 90 \\
\hline
Number of attention heads in multi-head attention & 4 & 5 \\
\hline
Dimension of the feedforward network inside each transformer layer & 16 & 16 \\
\hline
Number of transformer encoder layers & 2 & 2 \\
\hline
\end{tabular}
\end{minipage}
\hfill
\begin{minipage}{0.48\textwidth}
\centering
\caption{Hyperparameters for fault type classification and fault location identification}
\label{tab:hyperparameters}
\renewcommand{\arraystretch}{0.7}
\setlength{\tabcolsep}{3pt}
\tiny
\begin{tabular}{|p{4cm}|p{1.5cm}|p{1.5cm}|}
\hline
\textbf{Hyperparameters} & \textbf{Fault Type classification} & \textbf{Fault location identification} \\
\hline
Number of Attention Head & 4 & 5 \\
\hline
Optimizer & Adam & Adam \\
\hline
Learning rate & 0.001 & 0.001 \\
\hline
Batch Size & 32 & 32 \\
\hline
Epoch & 200 & 250 \\
\hline
Number of transformer encoder layers & 3 & 3 \\
\hline
Input feature dimension & 68 & 90 \\
\hline
Hidden dimension size in the feedforward layers inside the transformer encoder & 64 & 128 \\
\hline
Loss Function & Cross entropy loss & Cross entropy loss \\
\hline
\end{tabular}
\end{minipage}
\end{table}
\begin{figure}[H]
\centering
\begin{minipage}{0.55\textwidth}
\centering
\captionof{table}{Detailed layer-wise architecture of the FaultXFormer model for fault type and fault location identification.}
\label{tab:architecture}
\renewcommand{\arraystretch}{0.7}
\setlength{\tabcolsep}{3pt}
\tiny
\begin{tabular}{|l|c|c|}
\hline
\textbf{Layer} & \textbf{Output Shape} & \textbf{Parameters} \\
\hline
\multicolumn{3}{|c|}{\textbf{Fault Type Classification Model}} \\
\hline
Positional Encoding & [32,100,68] & 0 \\
Transformer Encoder Layer 1 & & 27,876 \\
-- MultiheadAttention (68→68) & [32,100,68] & 18,768 \\
-- Linear1 (68→64) & [32,100,64] & 4,416 \\
-- Linear2 (64→68) & [32,100,68] & 4,420 \\
-- LayerNorm1/2 & [32,100,68] & 272 \\
Transformer Encoder Layer 2 & [32,100,68] & 27,876 \\
Transformer Encoder Layer 3 & [32,100,68] & 27,876 \\
Global Avg Pooling & [32,68] & 0 \\
Linear (68→7) & [32,7] & 483 \\
\textbf{Total (Fault Type)} & — & \textbf{84,111} \\
\hline
\multicolumn{3}{|c|}{\textbf{Fault Location Identification Model}} \\
\hline
Positional Encoding & [32,100,90] & 0 \\
Transformer Encoder Layer 1 & & 56,378 \\
-- MultiheadAttention (90→90) & [32,100,90] & 32,760 \\
-- Linear1 (90→128) & [32,100,128] & 11,648 \\
-- Linear2 (128→90) & [32,100,90] & 11,610 \\
-- LayerNorm1/2 & [32,100,90] & 360 \\
Transformer Encoder Layer 2 & [32,100,90] & 56,378 \\
Transformer Encoder Layer 3 & [32,100,90] & 56,378 \\
Global Avg Pooling & [32,90] & 0 \\
Linear (90→20) & [32,20] & 1,820 \\
\textbf{Total (Fault Location)} & — & \textbf{170,954} \\
\hline
\end{tabular}
\end{minipage}
\hfill
\begin{minipage}{0.42\textwidth}
\centering
\includegraphics[width=\linewidth]{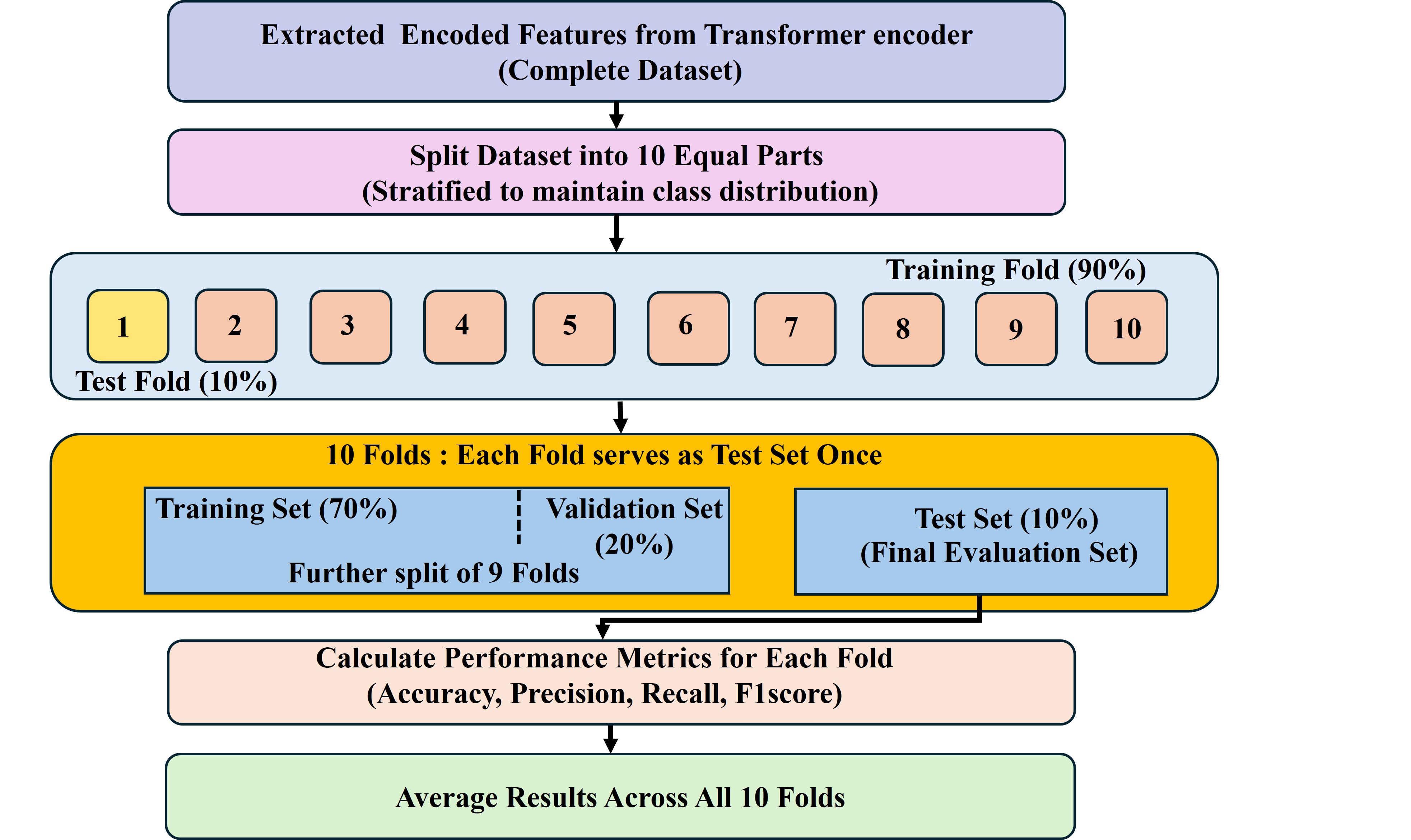}
\caption{Stratified 10-fold cross-validation of the FaultXformer model.}
\label{fig:10fold}
\end{minipage}
\end{figure}

\subsection{Stratified 10 Fold Cross Validation}
Figure \ref{fig:10fold} depicts a comprehensive 10-fold stratified cross-validation process for evaluating FaultXformer on fault analysis tasks. The approach begins with collecting encoded features from the transformer encoder using the whole dataset, which are then separated into 10 equal portions by stratified sampling to maintain class distribution balance. In each iteration, one fold (10\%) is used for testing, while the remaining nine folds (90\%) are divided into 70\% for training and 20\% for validation to optimize hyperparameters and prevent overfitting. Performance metrics are generated for each fold and averaged across all iterations to ensure a fair assessment of generalization.  As shown in Tables~\ref{table:results} and~\ref{tab:10fold_performance}, FaultXformer exhibits mean accuracies of 98.76\% for fault type classification and 98.92\% for fault location identification, with maximum accuracies of 99.56\% (Fold 5) and 99.74\% (Folds 4 and 8), illustrating consistent, robust, and generalizable performance across all data subsets.

The performance measurements showed minimal variation among folds, with coefficients of variation around 1\% for both fault type and fault location tasks, demonstrating strong stability and resilience. This consistency confirms that the model is mostly indifferent to training–testing partitioning. As stratified k-fold cross-validation presents an unbiased estimate of generalization error the assumption of identically and independently distributed (i.i.d.) samples \cite{john2010elements}, the near-invariant results empirically and theoretically validate the model’s reliability and generalization capability for fault identification and localization in medium-voltage distribution systems.

\begin{figure}[H]
    \centering
    \begin{minipage}{0.42\textwidth}
        \centering
        \includegraphics[width=\linewidth]{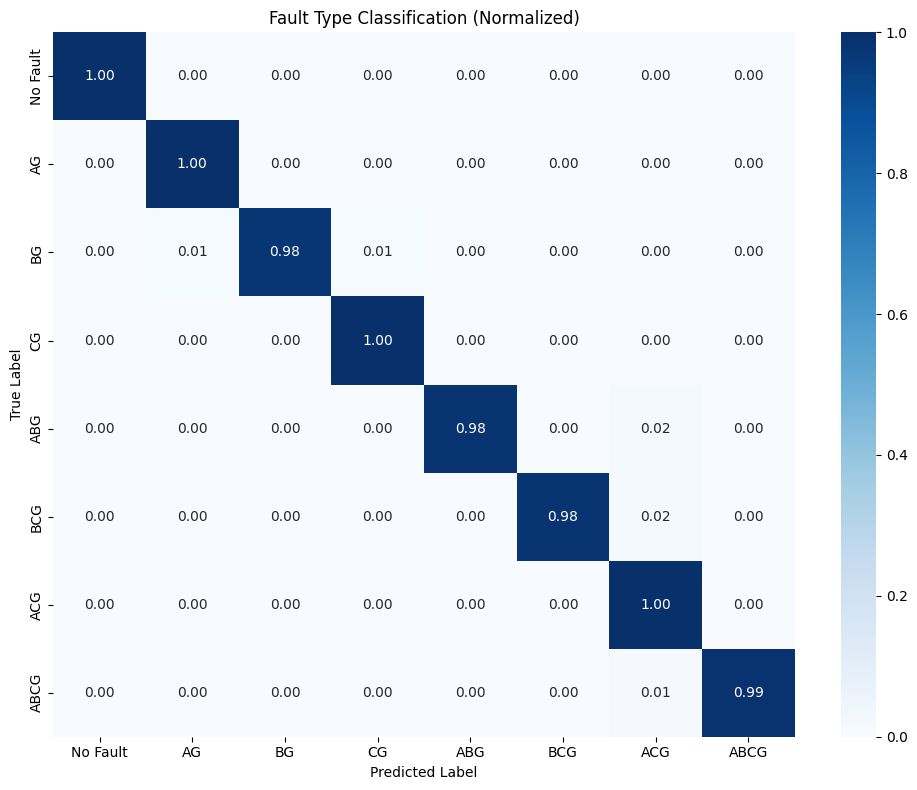}
        \caption{Confusion matrix for fault type identification.}
        \label{fig:cm_ft}
    \end{minipage}
    \hfill
    \begin{minipage}{0.42\textwidth}
        \centering
        \includegraphics[width=\linewidth]{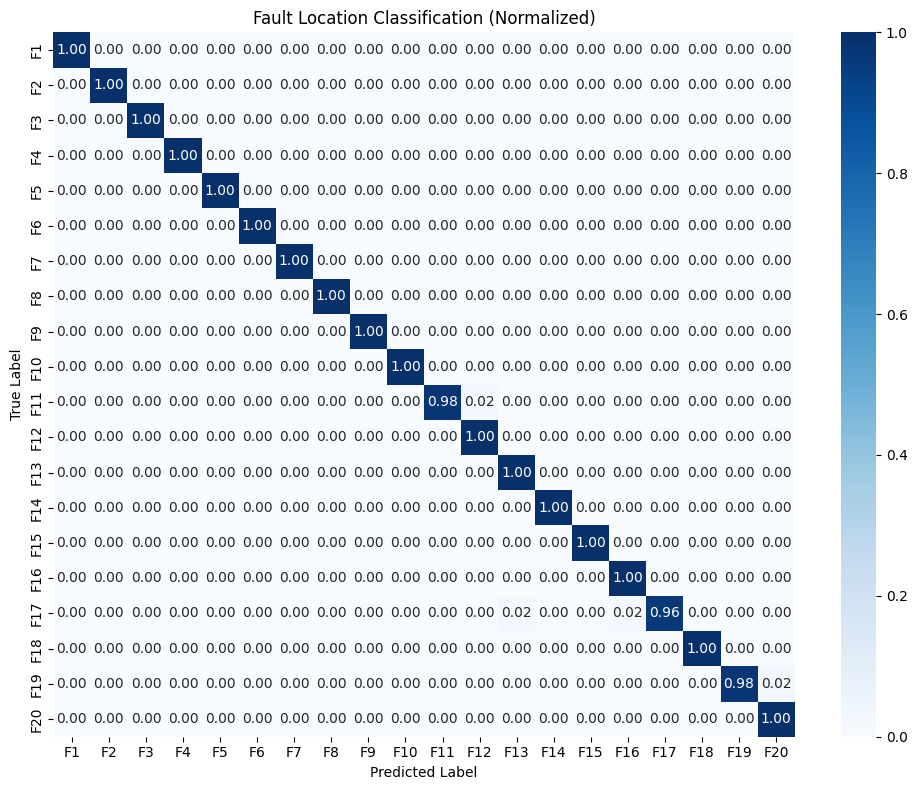}
        \caption{Confusion matrix for fault location identification.}
        \label{fig:cm_floc}
    \end{minipage}
\end{figure}
\begin{table}[H]
\centering
\begin{minipage}{0.48\textwidth}
\centering
\caption{Performance metrics across Stratified 10-fold cross-validation for fault type classification.}
\renewcommand{\arraystretch}{0.7}
\setlength{\tabcolsep}{3pt}
\tiny
\begin{tabular}{|c|c|c|c|c|}
\hline
\textbf{Fold} & \textbf{Accuracy} & \textbf{Precision} & \textbf{Recall} & \textbf{F1 Score} \\
\hline
Fold 1 & 0.9845 & 0.9840 & 0.9839 & 0.9839 \\
Fold 2 & 0.9933 & 0.9928 & 0.9927 & 0.9927 \\
Fold 3 & 0.9823 & 0.9833 & 0.9825 & 0.9826 \\
Fold 4 & 0.9823 & 0.9819 & 0.9835 & 0.9826 \\
Fold 5 & 0.9956 & 0.9952 & 0.9961 & 0.9956 \\
Fold 6 & 0.9911 & 0.9907 & 0.9903 & 0.9903 \\
Fold 7 & 0.9823 & 0.9821 & 0.9816 & 0.9815 \\
Fold 8 & 0.9845 & 0.9836 & 0.9850 & 0.9840 \\
Fold 9 & 0.9923 & 0.9928 & 0.9947 & 0.9937 \\
Fold 10 & 0.9867 & 0.9864 & 0.9853 & 0.9854 \\
\hline
\textbf{Average} & \textbf{0.9876} & \textbf{0.9873} & \textbf{0.9876} & \textbf{0.9872} \\
\hline
\end{tabular}
\label{table:results}
\end{minipage}
\hfill
\begin{minipage}{0.48\textwidth}
\centering
\caption{Performance metrics across Stratified 10-fold cross-validation for fault location identification.}
\renewcommand{\arraystretch}{0.7}
\setlength{\tabcolsep}{3pt}
\tiny
\begin{tabular}{|c|c|c|c|c|}
\hline
\textbf{Fold} & \textbf{Accuracy} & \textbf{Precision} & \textbf{Recall} & \textbf{F1 Score} \\
\hline
Fold 1 & 0.9936 & 0.9918 & 0.9949 & 0.9932 \\
Fold 2 & 0.9885 & 0.9847 & 0.9849 & 0.9834 \\
Fold 3 & 0.9949 & 0.9927 & 0.9962 & 0.9943 \\
Fold 4 & 0.9974 & 0.9981 & 0.9961 & 0.9971 \\
Fold 5 & 0.9898 & 0.9878 & 0.9911 & 0.9892 \\
Fold 6 & 0.9821 & 0.9874 & 0.9740 & 0.9789 \\
Fold 7 & 0.9770 & 0.9811 & 0.9777 & 0.9789 \\
Fold 8 & 0.9974 & 0.9975 & 0.9974 & 0.9974 \\
Fold 9 & 0.9898 & 0.9928 & 0.9873 & 0.9898 \\
Fold 10 & 0.9808 & 0.9850 & 0.9765 & 0.9792 \\
\hline
\textbf{Average} & \textbf{0.9892} & \textbf{0.9899} & \textbf{0.9876} & \textbf{0.9881} \\
\hline
\end{tabular}
\label{tab:10fold_performance}
\end{minipage}
\end{table}

\subsection{Class Wise Performance Evaluation}
A detailed class-wise evaluation for all fault types and locations presented in Tables~\ref{tab:fault_type_metrics} and~\ref{tab:fault_location_metrics} confirms the proposed model’s strong and consistent performance, exhibiting high accuracy, precision, recall, and F1 scores across classes.  The lowest accuracies were recorded for fault type BCG (98.15\%) and location F17 (95.92), however, overall results suggest strong generalization. The normalized confusion matrices (Figures~\ref{fig:cm_ft} and~\ref{fig:cm_floc}) reveal strong diagonal dominance, indicating great accuracy and low confusion, particularly among closely connected multi-phase faults and adjacent locations.
\begin{table}[H]
\centering
\begin{minipage}{0.48\textwidth}
\centering
\caption{Class-wise performance metrics for fault type classification.}
\label{tab:fault_type_metrics}
\renewcommand{\arraystretch}{0.7}
\setlength{\tabcolsep}{3pt}
\tiny
\begin{tabular}{|l|c|c|c|c|}
\hline
\textbf{Fault Type} & \textbf{Accuracy} & \textbf{Precision} & \textbf{Recall} & \textbf{F1 Score} \\
\hline
No Fault & 1.0000 & 1.0000 & 1.0000 & 1.0000 \\
AG       & 1.0000 & 0.9839 & 1.0000 & 0.9919 \\
BG       & 0.9894 & 1.0000 & 0.9894 & 0.9947 \\
CG       & 1.0000 & 1.0000 & 1.0000 & 1.0000 \\
ABG      & 0.9831 & 1.0000 & 0.9831 & 0.9915 \\
BCG      & 0.9815 & 0.9815 & 0.9815 & 0.9815 \\
ACG      & 1.0000 & 0.9846 & 1.0000 & 0.9922 \\
ABCG     & 0.9936 & 0.9924 & 0.9936 & 0.9934 \\
\hline
\end{tabular}
\vspace{0.2cm}
\tiny
\textbf{Lowest Performing Class:} BCG (Accuracy: 0.9815)
\end{minipage}
\hfill
\begin{minipage}{0.48\textwidth}
\centering
\caption{Class-wise performance metrics for fault location identification.}
\label{tab:fault_location_metrics}
\renewcommand{\arraystretch}{0.7}
\setlength{\tabcolsep}{3pt}
\tiny
\begin{tabular}{|l|c|c|c|c|}
\hline
\textbf{Fault Location} & \textbf{Accuracy} & \textbf{Precision} & \textbf{Recall} & \textbf{F1 Score} \\
\hline
F1  & 1.0000 & 1.0000 & 1.0000 & 1.0000 \\
F2  & 1.0000 & 1.0000 & 1.0000 & 1.0000 \\
F3  & 1.0000 & 1.0000 & 1.0000 & 1.0000 \\
F4  & 1.0000 & 1.0000 & 1.0000 & 1.0000 \\
F5  & 1.0000 & 1.0000 & 1.0000 & 1.0000 \\
F6  & 1.0000 & 1.0000 & 1.0000 & 1.0000 \\
F7  & 1.0000 & 1.0000 & 1.0000 & 1.0000 \\
F8  & 1.0000 & 1.0000 & 1.0000 & 1.0000 \\
F9  & 1.0000 & 1.0000 & 1.0000 & 1.0000 \\
F10 & 1.0000 & 1.0000 & 1.0000 & 1.0000 \\
F11 & 0.9778 & 1.0000 & 0.9778 & 0.9888 \\
F12 & 1.0000 & 0.9778 & 1.0000 & 0.9888 \\
F13 & 1.0000 & 0.9804 & 1.0000 & 0.9901 \\
F14 & 1.0000 & 1.0000 & 1.0000 & 1.0000 \\
F15 & 1.0000 & 1.0000 & 1.0000 & 1.0000 \\
F16 & 1.0000 & 0.9737 & 1.0000 & 0.9867 \\
F17 & 0.9592 & 1.0000 & 0.9592 & 0.9792 \\
F18 & 1.0000 & 1.0000 & 1.0000 & 1.0000 \\
F19 & 0.9792 & 1.0000 & 0.9792 & 0.9895 \\
F20 & 1.0000 & 0.9688 & 1.0000 & 0.9841 \\
\hline
\end{tabular}
\vspace{0.2cm}
\tiny
\textbf{Lowest Performing Fault Location:} F17 (Accuracy: 0.9592)
\end{minipage}
\end{table}

\subsection{Noise Sensitivity Analysis}
A noise sensitivity analysis was undertaken by adding Gaussian noise (1–3\%) into PMU current data to examine the Transformer encoder’s robustness under realistic settings.  The model maintained strong performance across all noise levels, obtaining peak accuracy of 0.9934 for fault type classification  at 2\% noise shown in Table \ref{tab:noise_analysisft} and an F1 score of 0.9560 for fault location identification at 3\% noise shown in Table \ref{tab:location_noise_analysis}.  These results validate the architecture’s remarkable resilience to signal distortion and its potential for reliable implementation in noisy, DER-integrated distribution systems.
\begin{table}[H]
\centering
\begin{minipage}{0.48\textwidth}
\centering
\caption{Noise sensitivity analysis for fault type classification. Performance metrics under increasing Gaussian noise levels.}
\label{tab:noise_analysisft}
\renewcommand{\arraystretch}{0.7}
\setlength{\tabcolsep}{3pt}
\tiny
\begin{tabular}{|c|c|c|c|c|}
\hline
\textbf{Noise Level (\%)} & \textbf{Accuracy} & \textbf{Precision} & \textbf{Recall} & \textbf{F1 Score} \\
\hline
1\% & 0.9912 & 0.9913 & 0.9911 & 0.9912 \\
2\% & 0.9934 & 0.9929 & 0.9934 & 0.9931 \\
3\% & 0.9889 & 0.9888 & 0.9884 & 0.9886 \\
\hline
\end{tabular}
\end{minipage}
\hfill
\begin{minipage}{0.48\textwidth}
\centering
\caption{Noise sensitivity analysis for fault location identification. Performance metrics at varying Gaussian noise levels.}
\label{tab:location_noise_analysis}
\renewcommand{\arraystretch}{0.7}
\setlength{\tabcolsep}{3pt}
\tiny
\begin{tabular}{|c|c|c|c|c|}
\hline
\textbf{Noise Level (\%)} & \textbf{Accuracy} & \textbf{Precision} & \textbf{Recall} & \textbf{F1 Score} \\
\hline
1\% & 0.9949 & 0.9950 & 0.9958 & 0.9954 \\
2\% & 0.9822 & 0.9793 & 0.9859 & 0.9820 \\
3\% & 0.9592 & 0.9548 & 0.9597 & 0.9560 \\
\hline
\end{tabular}
\end{minipage}
\end{table}

\subsection{Impact of DER Penetration}
A thorough evaluation was conducted to test the proposed framework’s resilience under varied DER penetration levels, ranging from 0\% to 80\%.  The models consistently exhibited high accuracy, exceeding 98\% for fault type classification and 97\% for fault location identification, exhibiting peak performances of 99.38\% and 99.62\% at 80\% DER, respectively as shown in Table \ref{tab:der_performance}. These results indicate the FaultXformer’s capacity to capture temporal–spatial dependencies in PMU data and maintain robustness under changing operating conditions, where increased DER penetration promotes signal variation and improves fault discrimination.
\begin{table}[H]
\centering
\caption{Performance of the proposed model under varying DER penetration levels.}
\label{tab:der_performance}
\resizebox{\columnwidth}{!}{%
\renewcommand{\arraystretch}{0.7}
\setlength{\tabcolsep}{3pt}
\tiny
\begin{tabular}{|c|c|c|}
\hline
\textbf{DER Penetration Level (\%)} & \textbf{Fault Type Accuracy (\%)} & \textbf{Fault Location Accuracy (\%)} \\
\hline
0  & 98.12 & 97.58 \\
20 & 98.83 & 97.45 \\
40 & 98.66 & 98.73 \\
60 & 99.07 & 99.36 \\
80 & 99.38 & 99.62 \\
\hline
\end{tabular}%
}
\end{table}

\subsection{Computational Efficiency and Real-Time Feasibility}
The real-time usability of the proposed FaultXformer model was assessed by measuring inference delay on both GPU and CPU platforms. According to the IEEE C37.118.1 standard\cite{power2014ieee}, the allowable end-to-end latency for PMU-based monitoring and protection applications typically ranges between 20 ms and 100 ms, based on the reporting rate and system requirement. The suggested model achieved an average inference latency of 17.35 ms per batch (0.54 ms per sample) on a CUDA-enabled GPU and 76.22 ms per batch (2.38 ms per sample) on a CPU.  Both configurations satisfy the IEEE real-time operational constraints, proving that the proposed framework is computationally efficient and suitable for online PMU data processing.  The GPU solution, in particular, delivers a fourfold reduction in latency, facilitating high-speed, scalable deployment for real-time fault detection and localization in medium-voltage networks.

\subsection{Interpretability Analysis Using Attention Visualization}

\begin{figure}[h]
\centering
\begin{minipage}{0.48\textwidth}
    \centering
    \includegraphics[width=6.5cm]{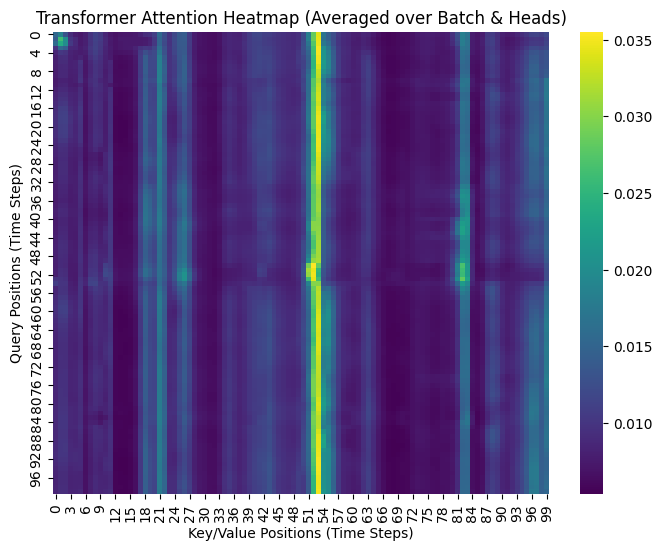}
    \caption{Visualization of temporal attention weights learned by the FaultXformer model.
    The heatmap shows how each time step (query) relates to others (key/value positions). The bright vertical band at time steps 51–54 highlights the model’s primary focus on the fault initiation phase, demonstrating interpretable and physically consistent attention behavior.}
    \label{fig:ah}
\end{minipage}
\hfill
\begin{minipage}{0.48\textwidth}
    \centering
    \captionof{table}{Accuracy of the FaultXformer model using different hyperparameters for fault type and fault location identification.}
    \label{tab:ablation}
    \renewcommand{\arraystretch}{0.7}
    \setlength{\tabcolsep}{3pt}
    \tiny
    \begin{tabular}{|l|c|c|c|}
    \hline
    \textbf{Parameter} & \textbf{Values} & \textbf{Fault Type (\%)} & \textbf{Fault Location (\%)} \\
    \hline
    Batch size & 16  & 95.15 & 94.67 \\
               & 32 & \textbf{99.33}& \textbf{99.74} \\
               & 64 & 96.42 & 95.89 \\
    \hline
    Latent dim & 32 & 88.37 & 96.28 \\
               & 68 & \textbf{99.33} & 96.42 \\
               & 90 & 96.82 & \textbf{99.74} \\
               & 100 & 96.27 & 97.63 \\
    \hline
    Layers & 1  & 89.27 & 85.61 \\
           & 2  & 93.74  & 91.22 \\
           & 3 & \textbf{99.33} & \textbf{99.74}\\
    \hline
    Heads & 2  & 95.68 & 93.27 \\
          & 4  & \textbf{99.33} & 95.66  \\
          & 5  & 97.63 & \textbf{99.74} \\
    \hline
    \end{tabular}
    \vspace{0.2cm}
    \tiny Values in bold indicate the best evaluation metrics for each task.

    \captionof{table}{Comparison of fault type classification and fault location identification accuracies using different methods.}
\label{tab:method_accuracy_comparison}
\renewcommand{\arraystretch}{0.7}
\setlength{\tabcolsep}{3pt}
\tiny
\begin{tabular}{|l|p{2cm}|p{2cm}|}
\hline
\textbf{Method} & \textbf{Fault Type Classification Accuracy (\%)} & \textbf{Fault Location Identification Accuracy (\%)} \\
\hline
RNN          & 64.38  & 58.85 \\
LSTM         & 97.29  & 93.47 \\
CNN          & 97.63  & 88.92 \\
FaultXformer (Proposed) & \textbf{99.33}  & \textbf{99.74} \\
\hline
\end{tabular}
\end{minipage}
\end{figure}

Although Transformer encoder architectures are often viewed as black boxes, the proposed framework achieves interpretability through its attention mechanism, as shown in Figure~\ref{fig:ah}, which illustrates the averaged attention heatmap over all heads and test samples, demonstrating the model’s temporal focus across phasor sequences. A clear concentration of attention around time steps 51–54 corresponds to the fault initiation interval, suggesting that FaultXformer consistently prioritises the most physically significant section of the signal. This illustrates that the model's judgements are based on physically significant electrical dynamics rather than arbitrary correlations, hence ensuring transparency, reliability, and interpretability in fault detection for power system applications.

\subsection{Ablation study on model parameters}
An ablation study was conducted to analyse the influence of major components and hyperparameters on FaultXformer’s performance. The initial investigations adopted baseline settings: learning rate 0.001, batch size 32, latent dimension 32, one Transformer encoder layer, and two attention heads. The following tests varied one parameter at a time. A batch size of 32 provided the highest accuracy for both fault type and location tasks. Optimal latent dimensions were 68 for fault type classification and 90 for fault location identification.  Increasing encoder layers to three consistently increased performance, while four and five attention heads produced the highest results for fault type and location identification, respectively.  These findings confirm the selected hyperparameters as best for each task.

\subsection{Comparison with Other Deep Learning Models}
A comparative analysis involving RNN, LSTM, and CNN under identical conditions (Table~\ref{tab:method_accuracy_comparison}) demonstrated that FaultXformer attained 99.33\% accuracy in fault type classification and 99.74\% in fault location identification, enhancing classification accuracy by 34.95\%, 2.04\%, and 1.70\%, and location accuracy by 40.89\%, 6.27\%, and 10.82\% compared to RNN, LSTM, and CNN, respectively. These results illustrate FaultXformer’s excellent capacity to capture long-range temporal and complicated spatiotemporal correlations in PMU data, ensuring robust and reliable fault diagnosis in active distribution networks.

\subsection{Limitations and Future Work}
Despite the promising results of the suggested FaultXformer model, numerous drawbacks should be addressed. First, the framework has been validated primarily on short-duration PMU sequences from the IEEE 13-node test feeder, restricting its shown application to this specific, small-scale distribution system. The model’s scalability to longer sequences and larger networks remains unproven, and its performance may require re-tuning and validation when applied to PMU datasets with varied noise characteristics or operational settings. Acknowledging these limitations provides important opportunities for future work to extend the model’s generalizability and robustness across diverse power system scenarios.

\section{Conclusion}
\label{section:conclusion}
This study developed FaultXformer, a Transformer-based framework for fault type classification and fault location identification in PMU-integrated active distribution networks.  The proposed approach takes advantage of a Transformer encoder to extract rich temporal-spatial information from raw PMU current signals (magnitude and phase), which are then given to two independently trained models optimized for their particular tasks.  Comprehensive evaluations were carried out utilising stratified 10-fold cross-validation.  The model achieved an average accuracy of 98.76\% for fault type classification and 98.92\% for fault location identification, exhibiting high precision, recall, and F1 scores across all folds.  Moreover, extensive classwise analysis indicated consistent performance across most fault categories, proving the model’s robustness in addressing varied fault scenarios.FaultXformer outperforms RNN, LSTM, and CNN models with enhancements of 34.95\%, 2.04\%, and 1.70\% in fault type classification accuracy, and 40.89\%, 6.27\%, and 10.82\% in fault location identification accuracy, respectively.

 In addition to cross-validation, a Gaussian noise sensitivity investigation (1\%–3\%) further confirmed the framework's robustness against realistic measurement noise, with both models maintaining good performance even under signal distortion.  This underlines the viability of FaultXformer for implementation in noisy, DG-integrated environments where conventional approaches generally fail.  The adaptability of the framework, particularly the employment of distinct classifiers for each task, allows for focused optimization and simpler scalability to new fault types or network conditions.

 Future studies could strengthen the existing framework by including voltage measurements, multi-modal fusion (e.g., SCADA + PMU), and real-time online learning to adapt to dynamic grid topologies.  Additionally, combining self-supervised pretraining techniques or temporal contrastive learning could further increase feature quality under limited or partially labeled data.  Deployment of FaultXformer on edge devices or within distributed control schemes also remains a potential avenue to provide proactive fault management in smart grids.

\singlespacing
\bibliography{main}

\end{document}